\documentclass{elsart}
\usepackage{graphics}
\usepackage{graphicx}
\usepackage{amssymb}
\usepackage{amsmath}

\begin{document}

\begin{frontmatter}

\title{Network Marketing on a Small-World Network}

\author{Beom Jun Kim \corauthref{cor1}}
\ead{beomjun@ajou.ac.kr}
\address{Department of Molecular Science and Technology, Ajou University,
Suwon 442-749, Korea}
\author{Tackseung Jun, }
\author{Jeong-Yoo Kim}
\address{Department of Economics, Kyung Hee University, Seoul 130-701, Korea}
\author{M. Y. Choi}
\address{Department of Physics, Seoul National University, Seoul
   151-747, Korea}
\address{ Korea Institute for Advanced Study, Seoul 130-722, Korea}
\corauth[cor1]{Corresponding author}

\begin{abstract}

We investigate a dynamic model of network marketing in 
a small-world network structure artificially constructed 
similarly to the Watts-Strogatz network model. 
Different from the traditional marketing, consumers can also 
play the role of the manufacturer's selling agents in
network marketing, which is stimulated by the referral fee
the manufacturer offers. As the wiring probability
$\alpha$ is increased from zero to unity, the network changes
from the one-dimensional regular directed network to the star network
where all but one player are connected to one consumer.
The price $p$ of the product and the referral fee $r$ are used as 
free parameters to maximize the profit of the manufacturer.
It is observed that at $\alpha=0$ the maximized profit is constant independent
of the network size $N$ while at $\alpha \neq 0$, it increases linearly
with $N$. This is in  parallel to the small-world transition.
It is also revealed that while the optimal 
value of $p$ stays at an almost constant level 
in a broad range of $\alpha$, that of  $r$ is sensitive
to a change in the network structure. The consumer surplus
is also studied and discussed.

\end{abstract}

\end{frontmatter}

\section{Introduction}
The traditional form of the market in economics can be best
described by the set of firms and consumers who sell and
buy goods. In a market, a trade between a consumer and a firm occurs 
if her valuation $v$
exceeds the product price $p$ quoted by the firm. This description,
however, hypothesizes that the product is known to consumers in 
the market, which is not always justified in real situations. 
For example, although a company produces a brand new
product which can potentially attract millions of people, the company
can sell very few if the information of this new
product has not been propagated yet across the public.
In a modernized society it costs a fortune only to
make the public know that there is a new product, and
companies are spending  tremendous amounts of money 
in massive advertisement on, e.g.,  a half-minute 
TV commercial or  a small portion of
a nationwide newspaper. Sponsoring Olympic games or World Cup football 
championship costs much more.

From the above reasoning, it is natural that some companies seek 
other ways to make the product information available without
spending the advertisement cost. One of the strategies along
this direction is to motivate a buyer to recommend the 
product to her social surroundings. An obvious way for
this is to pay the consumer if the referral induces the
actual purchase of the product by her social acquaintances. 
This is how network marketing works.

In this paper, we introduce a way of constructing a small-world
tree network~\footnote{In this paper, the term "small-world network" means
a network with a very short characteristic path length, irrespective
of the clustering property. In that sense, the tree network constructed here 
is a small-world network with a vanishing clustering coefficient.} 
 analogous to the Watts-Strogatz network
model~\cite{ref:network,WS} and use a game theoretic numerical approach
to simulate network marketing in which the product
is sold only through the connections in an existing
social network.  In parallel to the geometric
small-world transition that the average distance from the manufacturer 
ceases to increase linearly with the network size, it is observed
that manufacturer's profit exhibits striking difference
between the one-dimensional regular chain network and the
small-world tree network. We also investigate the consumer surplus
in network marketing, in comparison with conventional marketing.

The paper is organized as follows: 
In Sec.~\ref{sec:network}, the construction method of  
small-world tree networks is introduced. 
The game of consumer referrals is reviewed in Sec.~\ref{sec:game},
and extended to the case of general tree networks.
Section~\ref{sec:results} is devoted to the main result for
manufacturer's profit and also includes a discussion of the
consumer surplus. Finally, a brief summary is presented
in Sec.~\ref{sec:summary}.

\section{The Network}
\label{sec:network}

Social connections of people in societies have been studied in the
framework of complex network~\cite{ref:network}. In most
social networks, it is now well-known that they share 
characteristics in common such as the small-world behavior, a high level
of clustering, and so on. We here construct 
a directed small-world network in the same spirit as the Watts-Strogatz
model~\cite{WS} and then play the game of consumer referrals~\cite{jun}
on top of the network. 
To make the situation simple, we assign each consumer 
a unique value of the rank which is the 
distance (number of edges) from the company,
and assume that each consumer has only one precedent consumer except
the first consumer.  
A consumer at rank $R$ can refer all consumers 
at rank $R+1$ directly connected to her.
However, the reverse referral is forbidden (the product information
flows only in the downward direction from the company)
and accordingly, each consumer gets the referral only from one 
unique precedent consumer. In the graph theory, the above
structure is better captured by the directed tree graph.~\footnote{We
restrict ourselves to tree graphs to avoid the situation when
a consumer gets more than one referrals. See Sec.~\ref{sec:game}.}

\begin{figure}
\centering{\resizebox*{0.68\textwidth}{!}{\includegraphics{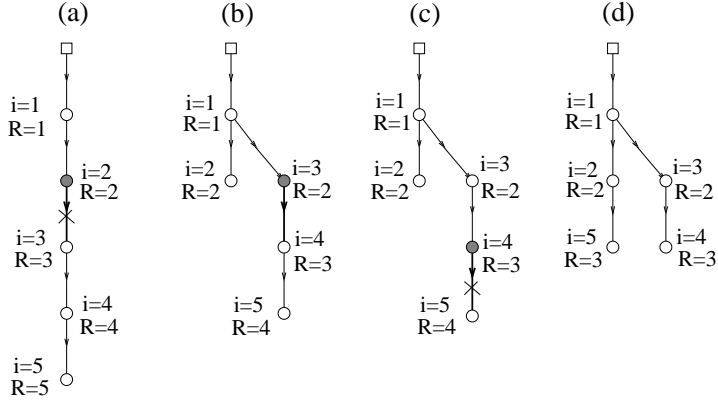}}}
\caption{Construction of the network of consumer referrals.
Starting from the one-dimensional chain in (a), each edge is
visited and then rewired with probability $\alpha$.
The edge $(j,j+1)$ with $j=2, 3$, and 4 in (a), (b), and (c), respectively,
is rewired to a randomly selected vertex
among the vertices whose ranks are less than the rank of $j$.
Open square is the manufacturer, and circles represent consumers.
At each step, the filled circle represents the vertex whose downward edge
is under consideration of rewiring. The crossed edge denotes
the situation when the rewiring is decided with probability $\alpha$.
The index $i$ of each vertex and its rank $R$, which is the distance
from the manufacturer, are shown for convenience.
See the text for more details.
}
\label{fig:network}
\end{figure}

Figure~\ref{fig:network} shows schematically the construction
method of the network used in this work. We first  build a
one-dimensional (1D) directed chain in Fig.~\ref{fig:network}(a), where
we have five consumers (circles) labelled as $i=1, 2, \cdots, 5$ 
and the company (square). The rank $R$ of the consumer $i$
in the initial 1D chain is thus given by $R_i = i$. Starting from the 
consumer $j=2$ [marked by a filled circle in Fig.~\ref{fig:network}(a)], we decide whether or not to rewire the edge $(j, j+1)$
with probability $\alpha$. If rewiring is decided [marked
by the crossed edge in Fig.~\ref{fig:network}(a)], all 
consumers and their connections below $j=2$ are attached to a
randomly selected vertex in the set $\{i | R_i < R_j\}$.
For example, in Fig.~\ref{fig:network}(a), $R_{j=2} = 2$ and 
there is only one vertex ($i=1$) with $R_i < 2$,
resulting in the network structure in Fig.~\ref{fig:network}(b).
The process goes on to $j=3$ in Fig.~\ref{fig:network}(b), which
shows the situation as an example that the edge is not selected to be rewired.
In Fig.~\ref{fig:network}(c) for $j=4$, the edge $(j,j+1)$ can
be rewired to either $i=1$ or $i=2$. In the case that there
are more than one vertex to which the edge can be rewired, one
is randomly picked: Shown in Fig.~\ref{fig:network}(d) is 
the situation that $i=2$ is selected. The whole process continues
subsequently from $j=2$ to $j=N-1$.

When the rewiring probability $\alpha = 0$, the network reduces
simply to a  one-dimensional regular chain network, while in the
opposite case of $\alpha = 1$, the network becomes
a star-like network in which vertices $i = 2, 3, \cdots , N$
are all connected to $i=1$. Similarly to the original Watts-Strogatz 
network~\cite{WS}, one can change the network structure
by varying the rewiring probability $\alpha$.

In most studies of complex networks, the characteristic
path length  defined as the average geodesic length connecting
each pair of vertices has been widely used to characterize
the structural property of the network. 
In particular, many networks have been shown to exhibit the so-called 
small-world behavior that the characteristic path length 
increases with the network size $N$ very slowly
(often in the logarithmic way). In the present study we define
the average rank $\langle R \rangle$ according to
\begin{equation}
\langle R \rangle\equiv \frac{1}{N}\sum_i R_i, 
\end{equation}
which is simply the number of edges from the manufacturer,
analogous to the characteristic path length in the
network literature.

\begin{figure}
\centering{\resizebox*{0.68\textwidth}{!}{\includegraphics{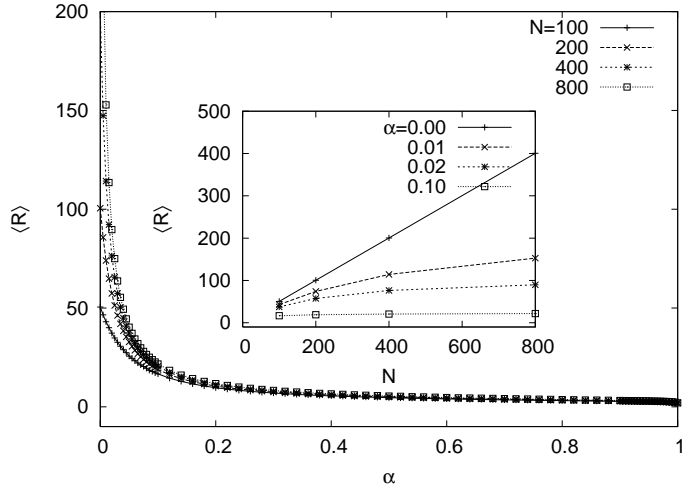}}}
\caption{The average rank $\langle R \rangle$ as a function of
the rewiring probability $\alpha$ for networks of sizes
$N=100, 200, 400,$ and 800. 
$\langle R \rangle$ is observed to decrease with $\alpha$.
   decays. Inset: $\langle R \rangle$ versus $N$
for $\alpha = 0$, 0.01, 0.02 and 0.1. As soon as $\alpha$ takes
a nonzero finite value, $\langle R \rangle$ ceases to increase
linearly with $N$, manifesting the small-world transition 
at $\alpha = 0$.
}
\label{fig:rank}
\end{figure}

In Fig.~\ref{fig:rank}, $\langle R \rangle$ is displayed as a function of
$\alpha$ for various sizes $N=100$, 200, 400, and 800. As $\alpha$ is
increased from zero, $\langle R \rangle$ reduces  monotonically,
becoming almost independent of $N$ for sufficiently large values of $\alpha$. 
The inset of Fig.~\ref{fig:rank} shows $\langle R \rangle$ as a function
of $N$ for $\alpha = 0$, 0.01, 0.02, and 0.10. It is clearly observed
that the cases $\alpha = 0$ and $\alpha \neq 0$ exhibit quite different behavior:
When $\alpha = 0$, $\langle R \rangle = (\sum_{i=1}^N i)/N = (N+1)/2 \propto N$,
while for $\alpha \neq 0$, $\langle R \rangle$ increases very slowly
with $N$. Consequently, the network of consumer referrals constructed
in the present work shows the small-world behavior at $\alpha \neq 0$,
similarly to the original Watts-Strogatz network~\cite{WS}.

\section{The Game of Consumer Referral}
\label{sec:game}
In this section, the game of consumer referrals is played on
the network constructed  in Sec.~\ref{sec:network}.
The manufacturer
produces the product at the marginal cost $c$ and sells it at
the price $p$ to consumers.
To make sense, $p$ should be larger than $c$ for  the company
tries to make a positive profit.
We assume that a consumer's valuation $v$ for the product
constitutes quenched random variables following given distribution function
$f(v)$. For simplicity,
we use the uniform probability distribution function:
$f(v) = 1$ for $v \in [0,1]$ and $f(v) = 0$ otherwise. 

The standard market works in a very simple way: The $i$th consumer 
has valuation $v_i$ and decides whether or not to buy according to 
the condition 
\begin{equation}
\label{eq:vimarket}
v_i > p.
\end{equation}
If the inequality is satisfied, the product is bought by the consumer
since she thinks that the product is worth her spending.
Accordingly, the probability of the purchase per consumer is simply
$1-p$ (for the uniform distribution of the valuation), 
yielding the total profit  of the manufacturer $\Pi$
\begin{equation}
\Pi = N(1-p)(p-c) - A, 
\end{equation}
where $A$ is the advertisement cost. 
The optimal price 
\begin{equation}
\label{eq:marketpmax}
p_{\rm max} \equiv (1+c)/2
\end{equation}
yields the maximum profit 
\begin{equation}
\label{eq:marketPimax}
\Pi_{\rm max} = \frac{N}{4} ( 1 - c)^2 - A .
\end{equation}

In the case of network marketing, the situation becomes more
complicated since each player can also make a profit if her referral induces
actual purchase. In our model, we include both the referral cost $\delta$ 
which is the cost to make one referral, and the referral fee $r$ which
is the initiative money the company pays in return for a successful referral.
Each consumer pays $\delta$ irrespective of the success of her
referral (e.g., the consumer has to make phone calls to persuade
her social surroundings to buy the product) while $r$ is paid to her 
on the condition of a successful referral.
As a simple example, consider the 1D regular network of two
consumers 1 and 2. The second consumer has none to refer and thus
her decision making is quite simple: She buys the product if $v_2 > p$.
On the other hand, the first consumer buys if 
\begin{equation}
\label{eq:v1}
v_1 > p - \max\{ r(1-p) - \delta , 0\}, 
\end{equation}
where $1-p$ is the probability that the second consumer buys the product.
When this happens the first consumer earns the money $r$. Regardless of the
success of her referral she should spend the cost $\delta$. 
If there is no
worth making referrals, i.e., when $r(1-p) - \delta$ is less than
zero, she simply does not make any referral but she still buys
if $v_1 > p$.~\footnote{
In economics, it is usually assumed that the participants in 
a market have {\em unbounded rationality}. In the present
consumer referral model, we also assume that each player is
smart, having full knowledge of the network
structure and understanding game dynamics completely.
From the physicist point of view, it is of interest to generalize
the model towards the case of bounded rationality~\cite{future}.} If there are $N$ consumers in the 1D regular
chain network, the purchase condition (\ref{eq:v1}) for consumer $i$ reads
\begin{equation}
v_i >  {\bar v}_i 
\end{equation}
with the minimum valuation
\begin{equation}
\label{eq:1dvi}
 {\bar v}_i = p - \max\{ r(1 - {\bar v}_{i+1}) - \delta, 0\},
\end{equation}
where $1-{\bar v}_{i+1}$ is the probability that the $(i+1)$th consumer
buys the product. Generalization of Eq.~(\ref{eq:1dvi}) to the network
structure in Sec.~\ref{sec:network} is straightforward:
\begin{equation}
\label{eq:networkvi}
 {\bar v}_i = p - \sum_{j \in \Lambda_i} \max\{ r(1 - {\bar v}_j) - \delta, 0\},
\end{equation}
where $\Lambda_i$ is the set of neighboring consumers of $i$ in the downward
direction from the manufacturer. For the consumers at the bottom ranks
the minimum  valuation is given by ${\bar v}_i = p$.

The consumer referral game in the present study goes as follows:
(i) The network of size $N$ with the rewiring probability $\alpha$
is built. (ii) The valuation $v_i$ of each consumer is assigned
from the uniform distribution in $[0,1]$. (iii) Starting from the
consumers at the end of the network whose ranks take the largest value,
we compute the minimum valuations as described above [${\bar v}_i = p$
for bottom consumers and Eq.~(\ref{eq:networkvi}) for others]. 
(iv) After the minimum valuation is computed  for every consumer,
we consider consumer $i$ at rank $R_i$, starting from $i=1$, and 
check the condition $v_i > {\bar v}_i$. If the inequality is fulfilled, 
the product is bought by $i$, and we proceed to the consumers at 
rank $R_i+1$. If the inequality is not satisfied and consumer $i$ 
does not buy the product, the chain of buyer's referral 
on the branch terminates. (v) The game ends when all chains of
buy-and-refer stop. When the game ends, we compute the profit of the
manufacturer 
\begin{equation}
\label{eq:networkPi}
\Pi = N_{\rm buyer}(p-c) - (N_{\rm buyer}-1)r,
\end{equation} 
where
$N_{\rm buyer}$ is the number of consumers who bought the product.
We fix $c$ and $\delta$ as constants and obtain the profit  $\Pi$
as a function of $p$ and $r$. From the point of view of the manufacturer,
we aim to optimize the profit with respect to both $p$ and $r$.

In our game of consumer referrals, we assume that only the consumers 
who actually bought the product make referrals. We believe that
this assumption is reasonable in view of that few people are willing 
to buy the product if the referrer herself has not bought it.

\section{Results}
\label{sec:results}
\subsection{Manufacturer's Profit}
\label{subsec:companyprofit}
Numerical simulations are performed at fixed values of the marginal
production cost $c = 0.05$ and the referral cost $\delta = 0.01$
(the use of other values, if not too different, is not expected
to change the main results). In the two-dimensional parameter space
$(p,r) \in [0, 1] \times [0,1]$, we first divide each parameter range
into ten intervals of the width $\Delta = 0.1$ and then compute the 
profit of manufacturer $\Pi(p,r)$ at the centers of 100 square boxes in 
the 2D parameter space. For the box where $\Pi$ has the largest value,
we repeat the above procedure four times with subsequently
narrower intervals ($\Delta = 0.01, 0.001,$ and $0.0001$).
When $\Pi(p,r)$ is computed for given values of $p$ and $r$,
we make an average over 10,000 different realizations of the network
structure and valuation.

\begin{figure}
\centering{\resizebox*{0.68\textwidth}{!}{\includegraphics{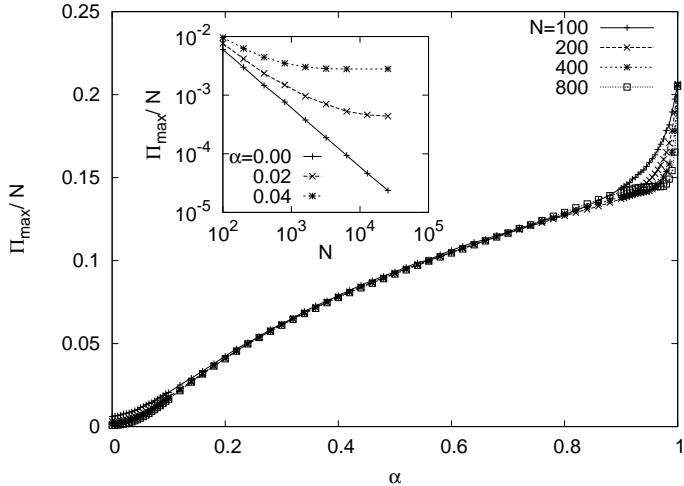}}}
\caption{The maximum profit of the manufacturer 
   per consumer $\Pi_{\rm max}/N$ versus the rewiring probability $\alpha$
   for networks of sizes $N=100, 200, 400,$ and $800$. 
      Inset: $\Pi_{\rm max}/N$  versus $N$ for $\alpha = 0.0, 0.02,$ and
         $0.04$.
}
\label{fig:profit}
\end{figure}

Figure~\ref{fig:profit} displays the maximum profit of the
manufacturer per consumer ($\Pi_{\rm max}/N$) as a function of 
the rewiring probability $\alpha$. As $\alpha$ is increased
from zero, $\Pi_{\rm max}$ is shown to be monotonically
increasing. Since all curves for different
network sizes overlap in a broad range of $\alpha$ except 
the region where $\alpha$ is close  to either zero or unity,
we henceforth focus on the system size $N=800$. One can also
conclude from Fig.~\ref{fig:profit} that the profit is 
proportional to the network size in the broad intermediate
region of $\alpha$ as expected.
In the inset of Fig.~\ref{fig:profit}, $\Pi_{\rm max}/N$ 
is exhibited as a function of $N$ at $\alpha = 0.0$, 0.02, and 0.04.
Clearly shown is that the behavior at $\alpha = 0$ is strikingly
different from those at $\alpha \neq 0$. Specifically,
for $\alpha = 0$ the maximum profit $\Pi_{\rm max}$
is constant irrespective of $N$
while it grows linearly with $N$ for sufficiently
large networks as soon as $\alpha$ takes a nonzero value.
This observation is in parallel to the small-world transition
discussed in Sec.~\ref{sec:network} (see Fig.~\ref{fig:rank}),
implying that the abrupt change in the behavior of
$\Pi_{\rm max}$ at $\alpha = 0 $ is closely related with the structural
small-world transition, reflected in the behavior
of $\langle R \rangle$.

The behavior around $\alpha = 1$ in Fig.~\ref{fig:profit}
is also interesting: As $N$ grows larger, the increase
of $\Pi_{\rm max}$ with $\alpha$ approaching unity appears to be
sharper, which indicates that in the thermodynamic limit 
($N \rightarrow \infty$)
$\Pi_{\rm max}$ should have discontinuity at $\alpha = 1$.
When $\alpha = 1$, the network reduces to a star network,
where all but one consumer ($i=2, 3, \cdots , N$) are connected to
the first consumer ($i=1$).   It is then straightforward to 
compute $\Pi_{\rm max}$ as follows: 
The minimum valuation for the $i$th ($i > 1$) consumer simply
reads $\bar v_i = p$ because she has none to refer, 
while  the first consumer has 
$\bar v_1 = p - (N-1)\max\{ r(1 - p) - \delta, 0 \}$.
The case of $ r(1-p) - \delta \leq  0$ is trivial: the first consumer
buys the product with probability $1-p$ but she does not make
any referral.
In the opposite case of $ r(1-p) - \delta > 0$, 
the minimum valuation for the first consumer becomes negative for 
sufficiently large
values of $N$ and consequently she always buys the product, 
resulting in the total number of buyers 
$N_{\rm buyer} = 1 + (N-1)(1-p)$. 
The manufacturer's profit per consumer is thus given by
$\Pi/N =  (1-p)(p-c-r)$
in the limit of large $N$, and the optimal value of $p$ is obtained 
from $\partial \Pi /\partial p = 0$, leading to the relation
$p_{\rm max} = (1 + c + r_{\rm max})/2$. Since the profit is a 
decreasing function of $r$, $r_{\rm max}$ is attained 
when $r_{\rm max}$ satisfies $r_{\rm max}(1-p_{\rm max}) = \delta$ 
and one finally gets 
$r_{\rm max} \approx 0.02$, $p_{\rm max} \approx 0.535$, and 
$\Pi_{\rm max}/N \approx 0.216$ for the values 
$c = 0.05$ and $\delta = 0.01$ used in this work.
This approximate value of $\Pi_{\rm max}/N$
is in reasonably good agreement with the value at $\alpha = 1$ in
Fig.~\ref{fig:profit}, apparently supporting
the existence of the discontinuity of $\Pi_{\rm max}/N$
at $\alpha = 1$ in the thermodynamic limit.

\begin{figure}
\centering{\resizebox*{0.68\textwidth}{!}{\includegraphics{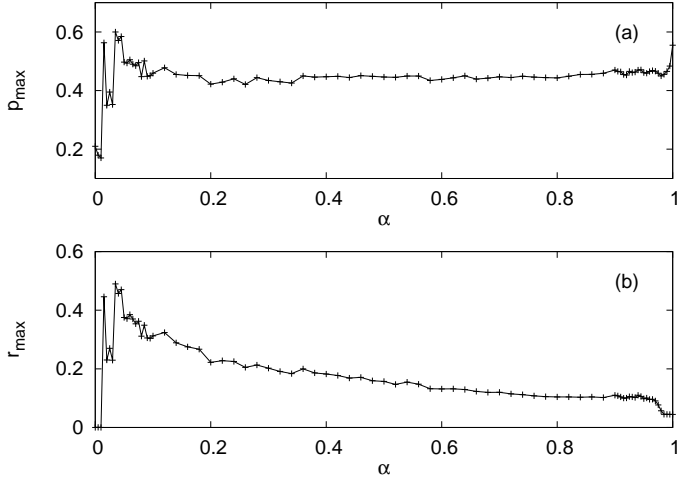}}}
\caption{The optimal  price $p_{\rm max}$ and the optimal referral fee
   $r_{\rm max}$ for the network of size $N=800$. 
}
\label{fig:pr}
\end{figure}

In Fig.~\ref{fig:pr}, the optimal values (a) $p_{\rm max}$
and (b) $r_{\rm max}$ are shown as functions of the rewiring
probability $\alpha$ for $N=800$. The values in Figs.~\ref{fig:profit}
and \ref{fig:pr} at $\alpha = 1$ are
reasonably well described by the above approximate calculations
[accurate only in the leading-order terms of $O(1/N)$].
It is very interesting to note that the optimal value of the
price $p_{\rm max}$ does not change significantly with $\alpha$ in 
a broad range of $\alpha$, while the optimal value of the referral fee
$r_{\rm max}$ keeps decreasing with $\alpha$.
Consequently, the above result suggests that in network marketing
the manufacturer needs to decrease the referral fee as the network
structure becomes more complex from the 1D chain network.

We next compare the above results obtained from network marketing
with the results for the conventional market.  The maximum profit
for the conventional market in Eq.~(\ref{eq:marketPimax}) is larger
for the corresponding value for network marketing shown in 
Fig.~\ref{fig:profit} when the advertisement cost $A$ is not so great. 
However, in case that $A$ grows large, e.g., for realistic values of $N$,
the manufacturer can make more profit by changing to network marketing.
For example, if the company spends 50\% of its profit as the
advertisement cost, the company can make more profit in network marketing
with $\alpha = 0.7$.  
\subsection{Consumer Surplus}
\label{subsec:consumerprofit}

\begin{figure}
\centering{\resizebox*{0.68\textwidth}{!}{\includegraphics{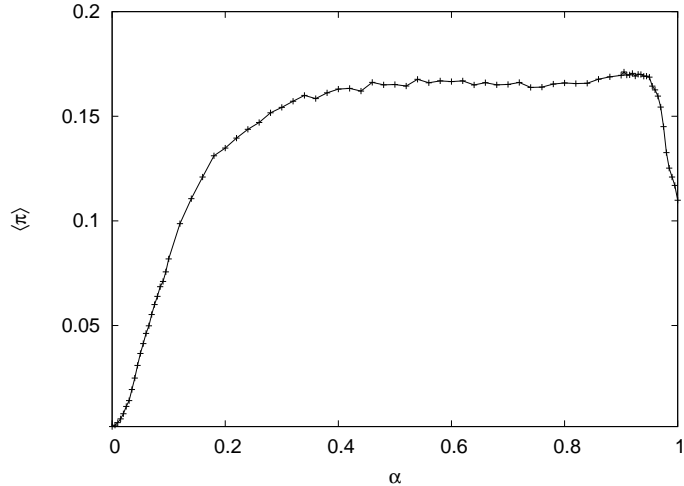}}}
\caption{The consumer surplus averaged over all consumers
   obtained at optimized values of $p_{\rm max}$ and $r_{\rm max}$
   (see Fig.~\ref{fig:pr}) for $N=800$ [see Eq.~(\ref{eq:consumerprofit})
   and the text for details].
}
\label{fig:consumerprofit}
\end{figure}

The monotonically increasing behavior of the maximum profit in 
Fig.~\ref{fig:profit} suggests that the manufacturer prefers
the situation in which all but one consumer buy without
referrals, and only one consumer makes all referrals and earns
a big  profit. This star-like network structure is simply
equivalent to the situation when the manufacturer makes a daughter 
firm which takes care of all issues related with
the distribution and the sale of the product.

We study the consumer surplus as follows: For given network
structure parameterized by $\alpha$, the manufacturer maximizes
its profit as described in Sec.~\ref{subsec:companyprofit}
by using the optimal values $p_{\rm max}$ and $r_{\rm max}$.
At given values of $p_{\rm max}$ and $r_{\rm max}$, the surplus 
for the $i$th consumer reads
\begin{equation}
\label{eq:consumerprofit}
\pi_i = r_{\rm max} n_i  - \delta k_i + (v_i - p_{\rm max}),
\end{equation}
where $k_i$ is the number of consumers (or  potential buyers) 
attached to $i$ , $n_i$ is the number of actual 
buyers among $k_i$ consumers,
and $v_i - p$ is the measure of $i$'s satisfaction in buying
the product (the larger $v_i - p$, the happier).
When the consumer $i$ does not buy the product her consumer
surplus $\pi = 0$ since she gets no referral fee from the
manufacturer and does not spend any referral cost.

Figure~\ref{fig:consumerprofit} displays the averaged consumer surplus
$\langle \pi \rangle \equiv (1/N)\sum \pi_i$.
Very interestingly, the consumer surplus
increases first, then stays almost at the same level
around $0.16$, and finally decreases again as $\alpha=1$ is approached.
It is thus concluded that while manufacturer's profit becomes the
highest at $\alpha = 1$, this does not lead to the
highest consumer surplus.

In the conventional market described by Eqs.~(\ref{eq:marketpmax})
and (\ref{eq:marketPimax}), the surplus $v - p$ is realized 
only when the consumer buys, i.e., when the inequality (\ref{eq:vimarket})
is satisfied, 
and accordingly the surplus per consumer
is given by
\begin{equation}
\label{eq:marketconsumer}
\langle \pi \rangle =  \int_0^1 \max( v - p, 0 ) f(v) dv
   = \int_p^1 (v - p) dv  , 
\end{equation}
where the uniform distribution $f(v) = 1$ for $v \in [0,1]$ has
been used.
Inserting Eq.~(\ref{eq:marketpmax}) to
Eq.~(\ref{eq:marketconsumer}), one obtains
\begin{equation}
\langle \pi \rangle \approx 0.11,
\end{equation}
where the same value of $c = 0.05$ has been used as in network
marketing. 
It is interesting to see that although conventional marketing
outperforms network marketing from the viewpoint of 
the manufacturer's profit (unless the advertisement cost is large),
the consumer surplus is much larger for network marketing
in a broad range of $\alpha$.

\section{Summary}
\label{sec:summary}
We have proposed a game theoretic way to study 
network marketing where consumers can also play the role
of selling agents motivated by the referral fee the company
offers in return for successful referrals. A simple model to build
directed tree networks has been introduced to investigate
the effects of the network structure on the game of network 
marketing. As the rewiring probability $\alpha$ is increased
from zero to unity, the network structure changes from a
one-dimensional regular chain  to a star network.
The manufacturer's profit is then numerically maximized by using
the two parameters in the game: the price of the product
and the referral fee. Observed is that the manufacturer's profit
takes the maximum value at $\alpha = 1$ (the star network) as
expected.
We  have also investigated the consumer surplus and found 
it higher at intermediate values of 
$\alpha$. Although the manufacturer's profit is higher in
conventional marketing than in network marketing in general,
the consumer surplus has been found to be opposite, i.e.,
higher in network marketing.
In real situations, the higher consumer surplus in network
marketing may motivate each consumer to broaden her 
social acquaintances, eventually increasing the total
number of potential buyers. When this happens, the manufacturer
can consider changing to network marketing.

\section*{Acknowledgments}
This work was begun in 1992 when the third author visited University at
Albany, SUNY and shaped when he delivered a seminar at KIAS.  He is very
grateful to Chong Kook Park and seminar participants at KIAS.
BJK was supported by grant  No. R01-2005-000-10199-0
from the Basic Research Program of the Korea Science
and Engineering Foundation.
Numerical computations have been
performed on the computer cluster Iceberg at Ajou University.


\begin{thebibliography}{30}
\bibitem{future}
More details of the present work including extensions to other more
realistic network structures with bounded rationality will be published
elsewhere.
\end{thebibliography}
\end{document}